\documentclass[12pt]{article}
\usepackage{amsmath}
\usepackage{bbm,amssymb}

\newcommand{\eq}[1]{\begin{equation}#1\end{equation}}

\newcommand{\al}[1]{\begin{align}#1\end{align}}
\newcommand{\subeq}[1]{\begin{subequations}#1\end{subequations}}

\def\cala         {{\cal A}}

\def\calf         {{\cal F}}

\def\calk         {{\cal K}}

\def\caln         {{\cal N}}

\def\calt         {{\cal T}}

\def\calw         {{\cal W}}
\def\calz         {{\cal Z}}

\def\Re           {{\rm Re\hskip0.1em}}
\def\Im           {{\rm Im\hskip0.1em}}

\begin{document}
\font\cmss=cmss10 \font\cmsss=cmss10 at 7pt
\leftline{\tt arXiv:0803.3149}

\vskip -0.6cm
\rightline{\small{\tt MPP-2008-21}}
\rightline{\small{\tt LMU-ASC 15/08}}

\vskip .7 cm

\hfill
\vspace{18pt}
\begin{center}
{\Large \textbf{Warped generalized geometry compactifications,\\ effective theories and non-perturbative effects}}
\end{center}

\vspace{6pt}
\begin{center}
{\large\textsl{Paul Koerber~$^{a}$ and Luca Martucci~$^{b}$}}

\vspace{25pt}
\textit{\small $^a$ Max-Planck-Institut f\"{u}r Physik -- Theorie,\\
                    F\"{o}hringer Ring 6,  D-80805 M\"{u}nchen, Germany}\\ \vspace{6pt}
\textit{\small $^b$ Arnold Sommerfeld Center for Theoretical Physics,\\ LMU M\"unchen,
Theresienstra\ss e 37, D-80333 M\"unchen, Germany}\\  \vspace{6pt}
\end{center}

\vspace{12pt}

\begin{center}
\textbf{Abstract}
\end{center}

\vspace{4pt} {\small Summarizing the results of \cite{effective}, we discuss
the four-dimensional effective approach to type II $N=1$ supersymmetric flux compactifications
with general $SU(3)\times SU(3)$-structure. In particular, we study the effect of a non-trivial
warp factor, which we argue leads naturally to a supergravity formulation invariant under local complexified
Weyl transformations. We show that the full ten-dimensional supersymmetry equations can be obtained as
F-flatness and D-flatness conditions from the superpotential and K\"ahler potential. We then consider non-perturbative corrections
to these supersymmetry conditions, following from adding instanton or gaugino condensation effects to the
superpotential. As examples, we show how smeared instantons allow to understand the ten-dimensional geometry of KKLT-like AdS vacua
and we give an explanation for the superpotential for ``mobile'' D3-branes in terms of a non-perturbatively
induced generalized complex structure.
\noindent }

\vspace{1cm}

\noindent {\em Contribution to the proceedings of the 3rd workshop of the RTN project `Constituents, Fundamental Forces and Symmetries of the Universe'
in Valencia, 1-5 October, 2007.}

\thispagestyle{empty}

\vfill
\vskip 5.mm
\hrule width 5.cm
\vskip 2.mm
{\small
\noindent e-mail: koerber at mppmu.mpg.de, luca.martucci at physik.uni-muenchen.de
}

\newpage
\setcounter{footnote}{0}

\section{$N=1$ flux backgrounds}

By now there is a huge literature on compactifications with fluxes (for reviews see \cite{fluxreviews})
as they feature potentials for some or even all of the scalars appearing in the four-dimensional
effective theory, and thus get rid of unobserved massless scalars, making way for phenomenologically
more interesting theories. Typically, some of the supersymmetry is assumed to be still present in
the four dimensions, while complete supersymmetry breaking is then to take place at a lower energy scale.
Without fluxes, the supersymmetry conditions lead to a Calabi-Yau manifold as compactification geometry,
and even in a special case of type IIB with fluxes it is possible to find warped Calabi-Yau geometry
as a solution. So far this is the most studied and best understood case.

In general however, it was shown in \cite{gmpt} that the supersymmetry conditions for $N=1$
supersymmetric flux compactifications of ten-dimensional type II supergravity
onto four-dimensional Minkowski or AdS$_4$ can be succinctly described in the language
of generalized complex geometry \cite{gencomplex}. Indeed, let us make the following compactification
ansatz for the ten-dimensional Majorana-Weyl supersymmetry generators
\eq{
\epsilon_1 = \zeta_+\otimes \eta^{(1)}_+ \, + \, \zeta_-\otimes \eta^{(1)}_- \ , \qquad
\epsilon_2 = \zeta_+\otimes \eta^{(2)}_\mp \, + \, \zeta_-\otimes \eta^{(2)}_\pm \ ,
}
where the upper/lower sign here and in the following is for type IIA/IIB respectively and $\zeta_-$ and $\eta^{(1,2)}_-$
are the complex conjugates of $\zeta_+$ and $\eta^{(1,2)}_+$.
$\zeta_+$ is an arbitrary four-dimensional
Killing spinor specifying the $N=1$ supersymmetry --- in AdS$_4$ it satisfies $\nabla_\mu \zeta_-=\pm \frac12 W_0 \gamma_\mu \zeta_+$
and in Minkowski just $\nabla_\mu \zeta_+=0$ --- while $\eta^{(1,2)}$ are fixed spinors characterizing
the geometry of the internal manifold $M$. Both $\eta^{(1)}$ and $\eta^{(2)}$ describe an $SU(3)$-structure so that the internal geometry is
said to possess an $SU(3)\times SU(3)$-structure. This naming convention can be slightly misleading since $\eta^{(1)}$ and $\eta^{(2)}$ are not
necessarily (everywhere) independent and likewise for the associated $SU(3)$s.
From $\eta^{(1,2)}$ one can build two polyforms of definite chirality, one of strictly even forms and
one of strictly odd forms. Appropriately normalized they read
\eq{
\Psi_\pm = - \frac{i}{||\eta^{(1)}||^2} \sum_l \frac{1}{l!} \, \eta^{(2)\dagger}_\pm \gamma_{m_1 \ldots m_l} \eta^{(1)}_+ dy^{m_l} \wedge \cdots \wedge dy^{m_1} \, .
}

As an example, $\eta^{(2)}=c \eta^{(1)}$ leads to the familiar case of strict $SU(3)$-structure and the above polyforms
become $\Psi_+=-i c^{-1} e^{iJ}$ and $\Psi_-=\Omega$, where $J$ and $\Omega$ are the (not necessarily closed) symplectic two-form
and holomorphic $(3,0)$-form. In the presence of O3 or O7-planes (and corresponding D-branes) for example, we have $c=\pm i$.

Renaming the polyforms as $\Psi_1 = \Psi_\mp$ and $\Psi_2 = \Psi_\pm$ for type IIA/IIB respectively, the supersymmetry conditions
in both theories are completely equivalent to the following differential conditions on $\Psi_{1,2}$
\subeq{\label{susycond}\al{
d_H \big(e^{4A-\Phi} \Re \Psi_1 \big) = & 3 \, e^{3A-\Phi} \Re (\bar{W}_0 \Psi_2) + e^{4A} \tilde{F} \, , \label{Fsusy1} \\
d_H \big[e^{3A-\Phi} (\bar{W}_0 \Psi_2)\big] =& 2 i \,|W_0|^2 e^{2A-\Phi}\Im \Psi_1 \, , \label{Fsusy2} \\
d_H(e^{2A-\Phi}\Im \Psi_1)= & 0 \ \label{Dsusy} ,
}}
where $d_H = d + H \wedge$ is the exterior derivative twisted by the NSNS three-form $H$, $A$ is the warp
factor, $\Phi$ the dilaton and the RR fluxes $F^{\text{tot}}$ split as follows in an external and an internal part
$F^{\text{tot}}=F + e^{4A} \text{vol}_4 \wedge \tilde{F}$.

There is a close relationship between these conditions and the
conditions for probes in these backgrounds to be supersymmetric. Indeed, it was found that
D-branes are supersymmetric iff they are generalized calibrated \cite{gencal,luca2,deforms,adsbranes} and the differential
conditions for the calibration forms of the different types of D-branes are exactly the above conditions.
In \cite{adsbranes} it was shown that this relation not only holds for Minkowski or AdS compactifications, but extends
to general static backgrounds.
For backreacting D-branes, it was found in \cite{paultsimpis} that the above background supersymmetry
conditions together with the source-corrected Bianchi identities of the form fields still imply the source-corrected
Einstein equation and dilaton equation of motion, provided that the D-brane sources are generalized calibrated.

\section{The four-dimensional effective action}

$N=1$ gauged supergravity is completely described by the K\"ahler potential, the superpotential and the gauge kinetic function.
First we will derive the superpotential from the tension of a domain wall, next we will argue that the most appropriate
formulation in the presence of a non-trivial warp factor is in fact the (partially gauge-fixed) superconformal action of \cite{supconf} and
finally we will construct the conformal K\"ahler potential. We will leave the analysis of the gauge kinetic function
for future work.

It turns out to be possible to organize the ten-dimensional fields into objects belonging to the four-dimensional
effective description without making a full reduction to a four-dimensional supergravity theory with a finite number of fields.
This full reduction, which would be the next step, would involve making a choice of suitable zero- or light modes in which to expand and would depend
on the details of the compactification \cite{suppot,red1,red2}. However, it turns out that we can already gain new insights in the current description,
for instance when adding a non-perturbative superpotential, a quantity usually associated to the four-dimensional description, as
we will demonstrate in section \ref{nonpert}.

\subsection{Superpotential from the domain wall tension}

The superpotential of the four-dimensional effective theory was already derived by ana\-ly\-sing
the gravitino supersymmetry variation or the gravitino mass in \cite{suppot}. Here we will
rederive it from a domain wall argument in a similar way as the Gukov-Vafa-Witten superpotential in
\cite{gvw} and obtain the correct dependence on the warp factor.

The tension of a BPS domain wall orthogonal to $x^3$ (spanning $\mathbb{R}$) and wrapping an internal generalized cycle $(\Sigma,\calf)$
is given by \cite{gencal}
\eq{
T_{\text{\tiny DW}}=2\pi\left| \int_\Sigma e^{3A-\Phi}\Psi_2|_\Sigma\wedge e^\calf\right|=
2\pi\left|\int_{\mathbb{R}\times M}\langle e^{3A-\Phi}\Psi_2, j^{\text{\tiny DW}}_{(\Sigma,\calf)}\rangle\right| \, ,
}
where $j^{\text{\tiny DW}}_{(\Sigma,\calf)}$ is the generalized current \cite{deforms} in $\mathbb{R}\times M$  associated to the domain wall. Using the Bianchi identities,  $d_H F=-j^{\text{\tiny DW}}_{(\Sigma,\calf)}$, we obtain
\eq{\label{branedw}
T_{\text{\tiny DW}}=2\pi\left| \int_{\{x^3=\infty\}\times M}\langle e^{3A-\Phi}\Psi_2, F\rangle-\int_{\{x^3=-\infty\}\times M}\langle e^{3A-\Phi}\Psi_2, F\rangle\right|\ .
}
From the four-dimensional point of view the D-brane can be seen as the domain wall separating two flux vacua. In \cite{cvetic}
it was argued that its tension is related to the difference between the values of the superpotential in the two configurations:
$T_{\text{\tiny DW}}=2|\Delta\calw|$. From this formula we can read off the {\em on-shell} superpotential, i.e.~the superpotential evaluated
in the vacuum
\eq{
\calw_{\text{on-shell}}=\pi\int_{M}\langle e^{3A-\Phi}\Psi_2, F\rangle\ .
}
To extrapolate the above on-shell expression to the complete off-shell superpotential we demand holomorphicity as a minimal requirement.
We should thus first find the natural complex variables of the setup.

A first natural holomorphic variable is suggested by the on-shell superpotential itself and is given by
\eq{
\label{holcoord1}
\calz \equiv e^{3A-\Phi}\Psi_2\ .
}
We can find the second holomorphic variable by looking at the action of a supersymmetric D-brane instanton, reading \cite{effective}
\eq{
S_{\text{E}}=2\pi\int_{\Sigma}(e^{-\Phi}\Re \Psi_1-i C)|_\Sigma\wedge e^\calf\ .
}
This suggests the following definition
for the second holomorphic coordinate
\eq{
\label{holcoord2}
\calt \equiv e^{-\Phi}\Re\Psi_1-i\Delta C\ ,
}
where we split the RR-fluxes as $F=F_0 + d \Delta C$ with $d F_0 = -j$,
and thus for the {\em off-shell} superpotential
\eq{
\calw = \pi\int_M\langle \calz,F_0+i\,d_H\calt\rangle
      = \pi\int_M\langle \calz,F+i\,d_H(\Re \calt)\rangle\ .
}

\subsection{The warp factor and the conformal K\"ahler potential}

The compactification ansatz for the metric
\eq{
\label{metricansatz}
d s^2=e^{2A(y)} g_{\mu\nu}(x) d x^\mu d x^\nu + h_{mn}(y) d y^m d y^n \ ,
}
is in fact invariant under a simultaneous shift of the warp factor and a rescaling of the metric
\eq{\label{weyl}
A\rightarrow A+\sigma \quad\text{and} \quad  g \rightarrow e^{-2\sigma}g \ .
}
Moreover, $\calz$ contains a redundancy in its phase so that the theory is also invariant under the chiral rotation
\eq{
\label{weylphase}
\calz\rightarrow e^{i\alpha}\calz\ .
}
The general $N=1$ supergravity with such a gauge invariance can be constructed by partially gauge-fixing the superconformal action of \cite{supconf}.
The conformal K\"ahler potential $\caln$ multiplies the Einstein-Hilbert term in the action of \cite{supconf}. From dimensional reduction
with the ansatz \eqref{metricansatz} we find
\eq{
\caln = 4\pi\int_M d^6y\sqrt{\det h}\, e^{2A-2\Phi} = \frac{\pi}{2}\Big(i\int_{M}e^{-4A}\langle\calz,\bar\calz\rangle\Big)^{1/3}\Big(i\int_{M}e^{2A}\langle t,\bar{t}\rangle\Big)^{2/3}\ ,
}
with $t=e^{-\Phi} \Psi_1$.

We can isolate the arbitrary complex factor in $\calz$
\eq{
\label{splitting}
\calz_{\text{old}} = Y^3 \calz_{\text{new}} \, ,
}
gauge-fix the complex Weyl invariance (\ref{weyl},\ref{weylphase}) --- we refer for more details to \cite{supconf} and \cite{effective} --- and obtain a standard supergravity in the Einstein frame
with
\subeq{\al{
\calw_{\text{\tiny E}} = & M_{\text{P}}^3\,\calw \, , \\
\calk = & - 3 \log \caln \, ,
}}
where $M_P$ is the four-dimensional Planck mass. In here, $\calw$ and $\caln$
have the same functional form, but the old $\calz$ is replaced with the new one defined in \eqref{splitting}.

\subsection{The supersymmetry conditions as F-flatness and D-flatness conditions}
\label{susyfromsup}

A careful analysis \cite{effective} (see also \cite{casbil}) shows that the F-flatness conditions
\eq{
\delta_{\calz,\calt} \calw - 3 (\delta_{\calz,\calt} \log \caln) \calw = 0 \, ,
}
where the variations are with respect to the holomorphic coordinates $\calz$
and $\calt$ defined in \eqref{holcoord1} and \eqref{holcoord2} respectively, exactly reproduce
the supersymmetry conditions \eqref{susycond} if $W_0 \neq 0$. When $W_0=0$ (in the Minkowski case)
eq.~\eqref{Dsusy} has to be added separately. This makes sense since it was also shown in \cite{effective}
that \eqref{Dsusy} can be interpreted as a D-flatness condition, which only when the vacuum expectation value
of the superpotential $W_0$ is not zero, follows on general grounds from the F-flatness condition.

This provides a non-trivial check on the superpotential $\calw$ and the conformal K\"ahler potential $\caln$.

\section{Non-perturbative corrections}
\label{nonpert}

Suppose we add a non-perturbative term $\calw_{\text{np}}$ to the superpotential generated by D-brane instantons or
stacks of space-filling D-branes undergoing gaugino condensation. It can be shown that this term
takes the form
\eq{
\calw_{\text{np}}=\cala\, \exp\Big( -\frac{2\pi}{n}\int_M \langle \calt ,\, j_{\text{np}}\rangle\Big) \, ,
}
where $j_{\text{np}}$ is the generalized current for the cycle wrapped by the instantons or the D-branes on which the gaugino condensation takes place,
 and the overall factor $\cala$ comes from the determinant of the Dirac action for the fermions on the D-branes,
 and should depend holomorphically on the background closed and open string degrees of freedom in a way consistent
 with the complexified Weyl invariance (\ref{weyl},\ref{weylphase}), which becomes the K\"ahler invariance in the Einstein frame. Its explicit form is generically hard to compute.

In the same way as outlined in section \ref{susyfromsup} one finds the non-perturbatively corrected supersymmetry
equations
\eq{
d_H\calz=\frac{2i(\calw +\calw_{\text{np}})}{\caln}\, e^{2A}\Im t+\frac{2i}{n}\,\calw_{\text{np}}\, j_{\text{np}}\ .
\label{susynonpert}
}

\subsection{KKLT-like AdS vacua from smeared instantons}

At tree level a supersymmetric type IIB compactification onto AdS$_4$ with strict $SU(3)$-structure
is impossible. In particular, it is not possible to solve \eqref{Fsusy2}. This seems in contradiction
with the solution of \cite{kklt}. However, that paper took into account non-perturbative corrections and
to account for their solution we should replace \eqref{Fsusy2} by \eqref{susynonpert}. By introducing instantons with
smeared current
\eq{
\tilde\jmath_{\text{np}} = \frac{\pi\sigma}{\caln}\,\, e^{2A}\Im t \ ,
}
where, in analogy with \cite{kklt}, we have defined
\eq{
\sigma=\int_{\Sigma}\Re\calt|_{\Sigma}\wedge e^\calf\ ,
}
it is possible to find a solution with $d_H \calz = 0$ and in the $SU(3)$-structure
case a solution with integrable complex structure. Indeed, substituting $\tilde\jmath_{\text{np}}$
in \eqref{susynonpert} and putting the right-hand side to zero we find
\eq{
\calw=-\calw_{\text{np}}\big(1+ \frac{\pi}{n}\sigma\big)\ .
}
Notice that there is an additional factor of $4/3$ multiplying $\sigma$ with
respect to eq.~(13) of \cite{kklt}, which comes from the failure of \cite{kklt}
to take into account the world-volume flux $\calf$ on the D-brane, which should
necessarily be non-zero since generically $0\neq H|_\Sigma=d \calf$.

\subsection{Mobile D3-branes}

Let us now consider the effect of a localized instanton. In the regime of reliability of the non-perturbative correction,
we expect the first term on the right-hand side of \eqref{susynonpert} to be small, as the estimate of the previous subsection confirms.
Thus, locally, we may consider the source
term as the leading one on the right-hand side of \eqref{susynonpert}, especially in the neighbourhood of the source, and write
\eq{\label{gcscorr2}
d_H\calz\simeq\frac{2i}{n}\,\calw_{\text{np}}\, j_{\text{np}}\ .
}
Let us now consider the warped Calabi-Yau IIB background for which classically $\calz=\Omega$.
In such a background probe D3-branes have classically a trivial superpotential, since \cite{luca2,deforms}
\eq{
d \calw_{D3}=-\pi \calz_{(1)}=0 \, .
\label{D3suppot}
}
However, in the presence of localized instantons, it follows from \eqref{gcscorr2} that $\calz$
develops a one-form part and thus deforms into a {\em generalized} complex structure
\eq{
d\calz_{(1)}=-\frac{2i}{n} \calw_{\text{np}}\,\delta^{(2)}(\Sigma)\ .
}
But we can easily see from \eqref{D3suppot} that in such a background D3-branes do feel a non-trivial superpotential.
For a small deformation $\calz_{(1)}$ and identifying $\calw_{\text{D}3}$ with ${\cal W}_{\text{np}}$
since $\calw_{\text{np}}$ should contain itself all the dependence on the space-filling D3 moduli, we
find
\eq{
\bar\partial(\partial \log \calw_{\text{np}})=\frac{2\pi i}{n}\,\delta^{(2)}(\Sigma)\ .
}
This is solved by
\eq{
\calw_{\text{np}} = f^{1/n}\tilde{\calw}_{\text{np}} \, ,
\label{d3sup}
}
where $\tilde{\calw}_{\text{np}}$ does not dependent on the D3-brane moduli and $f$ is the holomorphic section of the line bundle associated to the divisor $\Sigma$, which vanishes at the location of $\Sigma$ itself.
The final formula (\ref{d3sup}) is in perfect agreement with the results of \cite{ganor,haack,klebanov}, but gives a new ten-dimensional insight into  them. The key point is the deformation of the classical background complex structure into a non-perturbatively generated truly generalized complex structure, which is exactly what gives a non-trivial superpotential to the D3-branes in a geometric way along the lines of \cite{luca2,deforms}.

\section*{Acknowledgements}

P.K.~is supported by the German Research Foundation (DFG)
within the Emmy-Noether-Program (Grant number ZA 279/1-2); L.M.~ is supported by the DFG cluster
of excellence ``Origin and Structure of the Universe''.

\end{document}